\documentclass[aps,prb,twocolumn,superscriptaddress]{revtex4-1}
\usepackage{graphicx}% Include figure files
\usepackage{dcolumn}% Align table columns on decimal point
\usepackage{bm}% bold math

\usepackage{amssymb}

\begin{document}

%\preprint{}

\title{Carrier flow and nonequilibrium superconductivity \\ in superconductor-based light-emitting diode}


%\thanks{}

\author{Ryotaro Inoue}
\email{ryoinoue@crm.rcast.u-tokyo.ac.jp}
\affiliation{Research Center for Advanced Science and Technology, Univ. of Tokyo, Tokyo 153-8904, Japan.}%
\affiliation{CREST, Japan Science and Technology Agency, Kawaguchi 332-0012, Japan.}%
  
\author{Hideaki Takayanagi}
\affiliation{Dept. of Applied Physics, Tokyo Univ. of Science, Tokyo 162-8601, Japan.}%
\affiliation{MANA, National Inst. for Materials Science, Tsukuba 305-0044, Japan.}%
\affiliation{CREST, Japan Science and Technology Agency, Kawaguchi 332-0012, Japan.}%


\author{Tatsushi Akazaki}
\affiliation{NTT Basic Research Lab., Kanagawa 243-0198, Japan.}%
\affiliation{CREST, Japan Science and Technology Agency, Kawaguchi 332-0012, Japan.}%

\author{Kazunori Tanaka}
\affiliation{Central Research Lab., Hamamatsu Photonics K.K., Hamamatsu 434-8601, Japan.}
\affiliation{CREST, Japan Science and Technology Agency, Kawaguchi 332-0012, Japan.}%

\author{Hirotaka Sasakura}
\affiliation{Research Inst. for Electronic Science, Hokkaido Univ., Sapporo 060-8628, Japan.}%
\affiliation{CREST, Japan Science and Technology Agency, Kawaguchi 332-0012, Japan.}%

\author{Ikuo Suemune}
\affiliation{Research Inst. for Electronic Science, Hokkaido Univ., Sapporo 060-8628, Japan.}%
\affiliation{CREST, Japan Science and Technology Agency, Kawaguchi 332-0012, Japan.}%

\date{\today}

\begin{abstract}
Superconductor-based light-emitting diode (superconductor-based LED) in strong light-confinement regime are characterized as a superconductor-based three-terminal device, and its transport properties are quantitatively investigated.
In the gate-controlled region, we confirm the realization of new-type Josephson field effect transistor (JoFET) performance, where the channel cross-sectional area of the junction is directly modulated by the gate voltage.
In the current-injected region, the superconducting critical current of $\mu$A order in the Josephson junction is found to be modulated by the steady current injection of pA order.
This ultrahigh monitoring sensitivity of the radiative recombination process can be explained by taking into account the fact that the energy relaxation of the absorbed photons causes the conversion of superconducting pairs to quasiparticles in the active layer. 
Using quasiparticle density and superconducting pair density, we discuss the carrier flows together with the non-equilibrium superconductovity in the active layer and the superconducting electrodes, which take place for compensating the conversion. 
\end{abstract}

%\keywords{}

\maketitle

%\section{}
%\subsection{}
%\subsubsection{}

% If in two-column mode, this environment will change to single-column
% format so that long equations can be displayed. Use
%\begin{widetext}
%\end{widetext}

% figures should be put into the text as floats.
% Use the graphics or graphicx packages (distributed with LaTeX2e)
% and the \includegraphics macro defined in those packages.
% See the LaTeX Graphics Companion by Michel Goosens, Sebastian Rahtz,
% \begin{figure}
% \end{figure}
% that you will use with z\ref{} command in the braces of the \label{} command.
% entire page. There is no need to do explicit centering.

% Surround figure environment with turnpage environment for landscape
% figure
% \begin{figure}
% \end{figure}

\section{Introduction}

Superconductor-based light-emitting diode (superconductor-based LED) is expected to be the key device in quantum information technology because of its possible {\it on-demand} generation of entangled photon pairs.
In spite of the large mismatch in the energy scale, it is also predicted that the coherence of superconducting pair system can be transferred to photon systems \cite{AsanoPRL, JosephsonLED, JosephsonLaser}. 
The radiative recombination process in superconductor-based LED has been investigated by optical measurements, which revealed the enhanced oscillator strength \cite{HayashiAPEx}, high quantum efficiency and radiative recombination time rapidly decreasing with temperature \cite{SuemuneAPEx, SasakuraPRL, SuemuneJJAPReview}.
In transport measurements, on the other hand, the diffusion of superconducting pairs into the active layer was  demonstrated by DC and AC Josephson effect, and the monitoring sensitivity of the radiative recombination process was found to be several orders higher than that in optical measurements although the cause of this ultrahigh sensitivity has not been elucidated \cite{SuemuneAPEx, SasakuraPRL, InoueSuST}.
\par 
The structure of superconductor-based LED with one normal electrode at the p-type semiconductor side and two superconducting electrodes separated by a slit at the n-type semiconductor side can be considered as a superconductor-based three-terminal device.
The superconductor-based three terminal device, such as Josephson Field Effect Transistor (JoFET) \cite{AkazakiAPL}, supercurrent modulation device via normal carrier injection (including 0-$\pi$ transition) \cite{SchapersAPL, BaselmansNature}, have attracted considerable research interests form the viewpoints of not only superconducting electronics but also fundamental study of superconducting transport itself.
Especially in the case of superconductor-based LEDs with narrow slit width treated in this paper, most of the generated photons / photon pairs are absorbed immediately due to the strong light-confinement, and the transport of the device are strongly affected by the radiative recombination process.
In this paper, we quantitatively analyze the transport properties of superconductor-based LED characterized as a superconductor-based three-terminal device, and discuss the carrier flows and non-equilibrium superconductivity which are caused by the radiative recombination process.

\section{Experimental Setup}

\begin{figure}[htb]
\includegraphics[width=0.9\linewidth, clip]{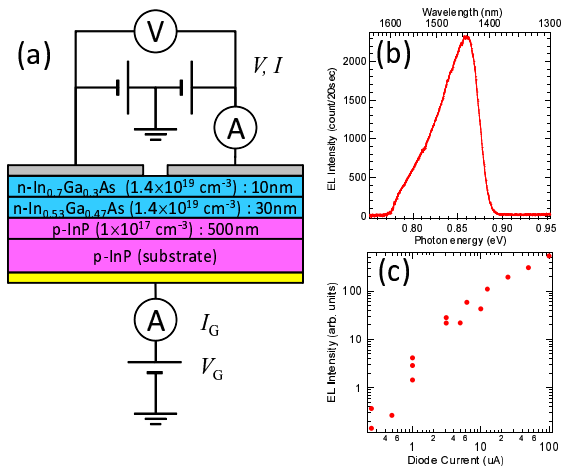}
\caption{(a) Schematic cross-sectional view of superconductor-based LED together with the measurement circuit. (b) EL spectrum of superconductor-based LED under the injection of 10 $\mu$A and (c) spectrally-integrated EL intensity as a function of injection. EL data are measured at 0.3 K, in the similar device with different gap between two Nb superconducting electrodes.}
\label{fig:Sample}
\end{figure}

Figure \ref{fig:Sample}(a) shows the schematic cross-sectional view of superconductor-based LED, where p-type indium phosphide (p-InP) layer and n-type indium gallium arsenide (n-InGaAs) layer are stacked on p-InP substrate, and form a p-n junction heterostructure.
When positive / negative voltage is applied to the gate electrode under the p-InP substrate, the p-n junction is biased forward / inversely.
Two niobium (Nb) superconducting electrodes with a thickness of 800 {\AA} are attached to the n-InGaAs layer, where superconducting pairs together with quasiparticles diffuse due to the proximity effect.
Under the forward-biased condition, normal holes injected from the p-InP layer recombine with the superconducting pairs and  quasiparticles in the n-InGaAs active layer.
The resultant electroluminescence (EL) emission can be detected from the slit between the two Nb electrodes, the width of which ($L$) is 150 nm.
Figures \ref{fig:Sample}(b) and \ref{fig:Sample}(c) show EL spectrum and spectrally-integrated EL intensity as a function of injected current, which were measured in the sample with the same composition but with different slit width.
It is found that the EL emission of $\sim$0.86 eV, which reflects the band structure of p-n junction, is obtained with the intensity roughly proportional to the injected current.
\par
The two Nb superconducting electrodes together with the n-InGaAs layer between them form a superconductor - semiconductor - superconductor Josephson junction structure.
We investigated the Josephson junction characteristics with changing the gate voltage ($V_G$) and / or the injected current ($I_G$), at a temperature of 30 mK using a dilution refrigerator. 
The measurement circuit is also shown in Fig. \ref{fig:Sample}(a).
In the measurement of Josephson junction characteristics, we biased the junction in such a way that the averaged electric potential of the two Nb electrodes was kept at 0 V. 
Therefore, when the hole current is injected from the gate electrode, the corresponding current is extracted via both of two Nb electrodes.
From the preliminary Hall measurement at 0.3 K, the carrier density and carrier mobility in the n-InGaAs layer were obtained as 1.4 $\times$ 10$^{19}$ cm$^{-3}$ and 1.6 $\times$ 10$^3$ cm$^2$/Vs, respectively, and we consider both of these values saturated in the low-temperature limit. 
Since the mean free path ($\ell$) and the thermal coherence length ($\xi_N$) at 30 mK are estimated to be $\sim$75 nm and $\sim$1.25 $\mu$m, the Josephson junction can be considered as {\it dirty} and {\it short} ($\ell < L < \xi_N$).

\section{Results and Discussion}

\begin{figure}[htb]
\includegraphics[width=0.8\linewidth, clip]{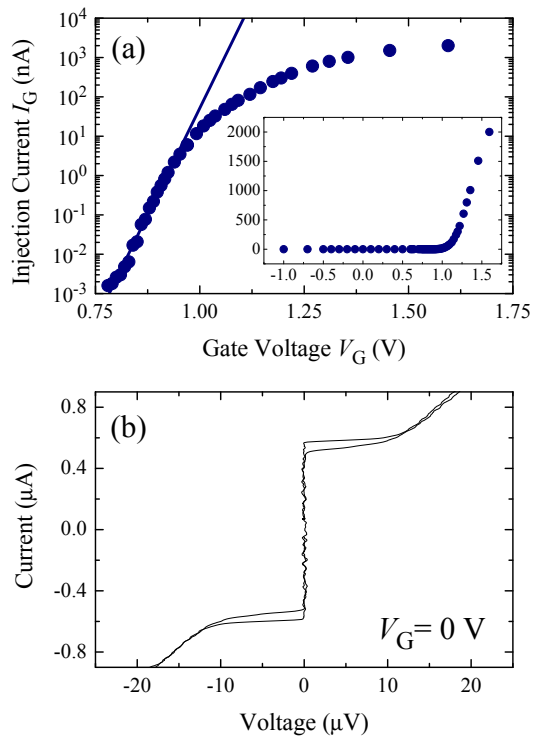}
\caption{(a)p-n junction characteristics of superconductor-based LED. The main panel and the inset are semilog plot and linear plot, respectively. (b) Josephson junction characteristics of superconductor-based LED with the gate voltage ($V_G$) of 0 V.}
\label{fig:Junction}
\end{figure}

\begin{figure}[htb]
\includegraphics[width=0.8\linewidth, clip]{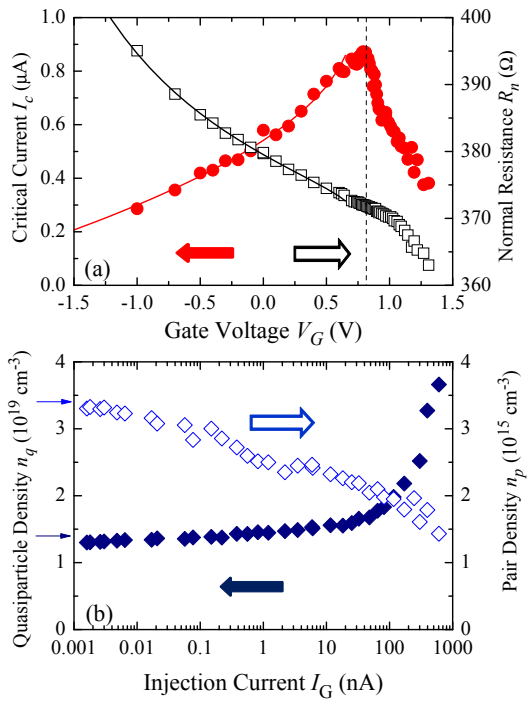}
\caption{(a)Superconducting critical current ($I_c$) and normal resistance ($R_n$) as functions of gate voltage ($V_G$). Fitting results in no gate-controlled region are represented by lines. (b) Estimated quasiparticle density ($n_q$) and pair density ($n_p$) in the active layer as functions of injected current ($I_G$). Arrows in the left vertical axis indicate the values of no injection limit.}
\label{fig:IcRnNqNp}
\end{figure}

Figure \ref{fig:Junction}(a) shows the injected hole current ($I_G$) as a function of applied gate voltage ($V_G$).
Steady current injection of $I_G =$ 1.6 pA starts at $V_G =$ 0.78 V.
Although we obtain the typical p-n junction diode characteristics $I_G \sim \exp (eV_G / \eta k_B T)$ with a non-ideal coefficient $\eta \sim$ 7.7 $\times$10$^3$, the tendency of saturation can be observed at $I_G \gtrsim$ 10 nA.
The hysteretic current - voltage characteristics of the Josephson junction with $V_G =$ 0 V are shown in Fig. \ref{fig:Junction}(b). 
The superconducting critical current ($I_c$) and the normal resistance ($R_n$) defined in sufficiently large current range, were 0.58 $\mu$A and 380 $\Omega$, respectively.
With the increasing temperature upto 160 mK, $I_c$ declined as $\sim T^{-1/4}$ while $R_n$ took a constant value.
When we applied the gate voltage ($V_G$), both $I_c$ and $R_n$ were found to be changed as shown in Fig. \ref{fig:IcRnNqNp}(a), which we will discuss in the following.
 
In the gate-controlled region of $V_G <$ 0.78 V, where we do not observe steady current injection, a depletion layer remains at the boundary of p-InP - n-InGaAs heterostructure.
In this region, we succeeded the quantitative explanation of the behavior of Josephson junction characteristics ($I_c$ and $R_n$) by taking into account the depletion layer thickness modulated by the gate voltage ($V_G$).
The critical current ($I_c$) and the normal resistance ($R_n$) can be written as follows in {\it dirty} and {\it short} junctions using the effective mass ($m^\ast$) and mobility ($\mu$) of carrier \cite{Tinkham}:
\begin{eqnarray}
I_c &=& \frac{e n_p \hbar}{m^\ast} \frac{W (H-x(V_G))}{L} , \nonumber \\
R_n &=& R_0 + \frac{1}{en_q\mu} \frac{L}{W (H-x(V_G))} .
\label{eqn:IcRn}
\end{eqnarray}
\noindent
Here, $L$ (150 nm), $H$ (40 nm) and $W$ are the length, thickness and width of the junction, and $n_q$ and $n_p$ are the quasiparticle density and superconducting pair density, respectively. 
As for the depletion layer thickness at the boundary of heterostructure ($x(V_G)$), we assumed the standard functional form of $x(V_G)=x(0)(1-V_G/V_i)^{1/2}$.
By the fitting shown in Fig. \ref{fig:IcRnNqNp}(a) and the values of carrier density and carrier mobility in the n-InGaAs layer estimated from the Hall measurement, we obtained $W$=1.02 $\mu$m, $R_0$=360.8 $\Omega$, $x(0)$=17.8 nm and $V_i$=0.7 V.  
The effective width of junction $W$ (1.02 $\mu$m) is significantly smaller than the geometrical width of junction $W_0$ (20 $\mu$m), which implies that the supercurrent flows only through narrow paths in the n-InGaAs layer.
Using the interfacial resistance ($R_0$), we defined the channel resistance ($R_{\rm Ch}$) as $R_{\rm Ch} \equiv R_n - R_0$ and investigated the $I_c R_{\rm Ch}$ product, which is expected to be independent of junction dimensions ($L$, $W$ and $H-x(V_G)$) and proportional to the ratio of carrier densities ($I_c R_{\rm Ch} = (\hbar / m^\ast \mu) (n_p / n_q)$) from Eq. (\ref{eqn:IcRn}).
The obtained $I_c R_{\rm Ch}$ product took a constant value of 10 $\mu$V in the gate-controlled region, which quantitatively supports our description of the direct modulation of channel cross-sectional area by the gate voltage.
This description can be regarded as a new-type JoFET, and we emphasize the difference from the conventional JoFET \cite{AkazakiAPL} where the carrier density in semiconductor (normal conductor) is modulated by the gate voltage.
The estimated value of superconducting pair density $n_p$ (3.4 $\times$ 10$^{15}$ cm$^{-3}$) is comparable to those estimated in other superconductor-based LEDs \cite{SuemuneAPEx}. 
\par
When the steady current injection takes place in the current-injected region of $V_G \geq$ 0.78 V, both of $I_c$ and $R_n$ decrease.
We note that the magnitude of injected current ($I_G$) is several orders smaller than the modulated superconducting critical current ($I_c$).
Therefore, the mechanism of this ultrahigh monitoring sensitivity is substantially different from those in conventional carrier-injected devices \cite{SchapersAPL, BaselmansNature}, and can be considered to reflect the radiative recombination process, which characterizes the superconductor-based LEDs with strong light-confinement. 
In fact, we cannot explain the behaviors of Josephson junction characteristics ($I_c$ and $R_n$) in small current injection region of $I_G \lesssim$ 10 nA by the increase of effective temperature of quasiparticle - pair system in the n-InGaAs active layer because $R_n$ takes constant value in the temperature range less than 160 mK.
If we take into account the fact that the depletion layer does not exist in this region ($x(V_G) = 0$) and reconsider Eq. (\ref{eqn:IcRn}), the carrier density in the n-InGaAs active layer ($n_q$ and $n_p$) are found to be the only variables that can explain the behaviors of $R_n$ and $I_c$.
Figure \ref{fig:IcRnNqNp}(b) shows $n_q$ and $n_p$ estimated from Eq. (\ref{eqn:IcRn}) as functions $I_G$.
\par
First, it is noteworthy that the sum $(n_q + 2n_p)$ increases with $I_G$ in all current-injected region.
Considering the charge neutrality condition, this increase indicates that the energy of conduction band for electrons in the n-InGaAs active layer is pulled down with respect to vacuum by the current injection.
This also means that the energy of valence band for holes is pulled up, which is implied by the tendency of saturation shown in Fig. \ref{fig:Junction}(a).
We also note that both $n_q$ and $n_p$ do not show the tendency toward saturation in the limit of $I_G \rightarrow 0$, the values of which are indicated by arrows in the left vertical axis of Fig. \ref{fig:IcRnNqNp}(b).

\begin{figure}[htb]
\includegraphics[width=0.7\linewidth, clip]{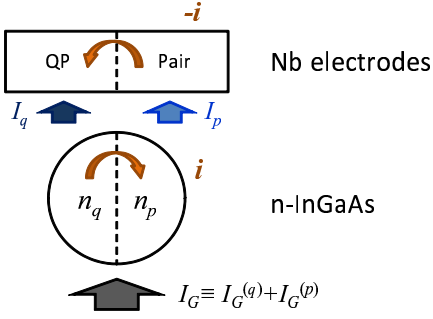}
\caption{Schematic diagram of quasiparticles and superconducting pairs.}
\label{fig:Cycle}
\end{figure}

\par
Using the flow diagram of quasiparticles and superconducting pairs schematically shown in Fig. \ref{fig:Cycle}, we discuss the carrier flow in the superconductor-based LED and explain the mechanism of the ultrahigh monitoring sensitivity for radiative recombination process.
Because the total charge conserves in the energy relaxation process in succession to the photon absorption, current with an amount exactly equal to $I_G$ is required to be extracted from the n-InGaAs active layer to the Nb electrodes in the steady state.
This means that the carriers flow into the conduction band of the n-InGaAs layer to compensate the carrier loss caused by the recombination process.
We define $I_q$ and $I_p$ as the current carried by quasiparticles and pairs, and represent the steady state condition as 
\begin{equation}
I_q + I_p = I_G \equiv I_G^{(q)} + I_G^{(p)}.
\label{eqn:neutrality}
\end{equation}
\noindent
Here $I_G^{(q)}$ and $I_G^{(p)}$ are the components of injected hole current ($I_G$) that recombine with quasiparticles and pairs corresponding to their recombination rates.
\par
In addition to the inflow of carriers ($I_q$ and $I_p$), the hole current injection causes large disturbance in the quasiparticle - pair system via radiative recombination and immediate absorption of generated photons, and promotes the conversion from pairs to quasiparticles.
If we put the current corresponding to the conversion rate as $i$, the following condition is also required in the steady state: 
\begin{equation}
0 = I_G^{(q)} - I_q - i = I_G^{(p)} - I_p + i.
\label{eqn:steady}
\end{equation}
\noindent
(We define the positive direction of $i$ as the flowing direction from quasiparticle subsystem to pair subsystem taking into account that the carrier charge is negative.) 
\par
Generally speaking, the conversion rate from quasiparticles to pairs ($i$) depends not only on the injected current ($I_G$) but also on the carrier density ($n_q$ and $n_p$).
However, in the small current injection region of $I_G \lesssim$  10 nA, we can assume that the energy relaxation process of absorbed photons dominantly proceeds with the destruction of superconducting pairs and put $i \approx \alpha I_G$.
Here the proportional constant $\alpha$ is approximately equal to the ratio of the photon energy ($\hbar \omega_{\rm op} \sim$ 0.86 eV) to the superconducting pairing energy (${\tilde \Delta}_N \sim$ 1.5 meV in Nb), i.e. $\alpha \approx \hbar \omega_{\rm op} / {\tilde \Delta}_N$.
After solving Eqs. (\ref{eqn:neutrality}) and (\ref{eqn:steady}) with respect to $I_q$ and $I_p$, and taking into account that $\alpha \gg 1$, we obtain 
\begin{eqnarray}
I_q &=& -(\alpha -1) I_G^{(q)} - \alpha I_G^{(p)} \approx  - \alpha I_G , \nonumber \\
I_p &=& \alpha I_G^{(q)} + (\alpha +1) I_G^{(p)} \approx \alpha I_G .
\end{eqnarray}
\noindent
That is, in order to compensate the conversion from pairs to quasiparticles inside the n-InGaAs active layer, the superconducting pairs flow {\it into} the n-InGaAs layer ($I_p>0$) while the quasiparticles flow {\it out of} ($I_q<0$).
(To be accurate, the outflow of quasiparticles is a little larger than the inflow of pairs and the steady state condition is secured including the injected hole current.)
Both $\vert I_q \vert$ and $\vert I_p \vert$ are larger than $I_G$ by the factor of $\alpha \gg 1$, which results in the high sensitivity of Josephson characteristics for monitoring the radiative recombination process.
\par
The values of $n_q$ and $n_p$ in steady state are determined as the balance points in the carrier flow.
Because the superconducting pair potential is sufficiently small, the reformation of pairs inside the n-InGaAs layer is negligible and the pairs lost in the energy relaxation process are compensated only by the inflow of pairs from Nb electrodes ($I_p$).
In the Nb superconducting electrodes, the discrepancy of chemical potentials in quasiparticle - pair system takes place in order to sustain the reformation rate necessary for the compensation, which also means the nonequilibrium superconductivity in the Nb electrodes.
Due to the large interfacial barriers at Nb electrodes ($R_0$), it is difficult to estimate the carrier density in the Nb electrodes ($n_q^{\rm (Nb)}$ and $n_p^{\rm (Nb)}$) directly from those in the n-InGaAs active layer ($n_q$ and $n_p$).
However, we can say that the sudden decrease of $n_p$ in the region of $I_G \lesssim$ 10 nA reflects the decrease of $n_p^{\rm (Nb)}$ caused by the rate-limiting reformation.
\par
When the injected current is increased ($I_G \gtrsim$ 10 nA), we cannot ignore the recombination process involving quasiparticles in addition to various energy relaxation processes of absorbed photons other than the destruction of superconducting pairs.
Moreover, the effective temperature of quasiparticle - pair system in the n-InGaAs active layer is increased, which causes not only the damping of the compensation cycle but also the suppression of $I_c$ itself.
These effects result in the gradual decrease of the current amplification factor ($\alpha$), and eventual break down of the flow model described in Fig. \ref{fig:Cycle}.
Especially in the region of $I_G \gtrsim$ 0.3 $\mu$A, the transport measurements of Josephson junction characteristics under the steady current injection become difficult whereas the spectrally-integrated EL intensity reaches the sensitivity of our optical measurement system as is shown in Fig. \ref{fig:Sample}(c).
The injected current ($I_G$) comparable with the superconducting critical current ($I_c$) causes asymmetric distortion of current - voltage curve which is due to the small difference in the transparency of interfaces at the two Nb electrodes.

\section{Conclusion}

In conclusion, we characterized superconductor-based LED in strong light-confinement regime as a superconductor-based three-terminal device, and quantitatively investigated its transport properties ($I_c$ and $R_n$).
In the gate-controlled region, we confirmed the realization of new-type JoFET performance, where the channel cross-sectional area of the junction is directly modulated by the gate voltage.
In the current-injected region, the superconducting critical curren of $\mu$A order was found to be modulated by the steady current injection of pA order.
We explained this ultrahigh monitoring sensitivity for radiative recombination process by taking into account the conversion of superconducting pairs to quasiparticles in the energy relaxation of absorbed photons, and discuss the carrier flows together with the nonequilibrium superconductivity, both of which take place for compensating the conversion. 

%\appendix
%\section{}


\begin{acknowledgments}
We appreciate Dr. S. Kim for fruitful discussion.
\end{acknowledgments}


\end{document}